# Coupled elasticity in soft solid foams

F. Gorlier[1], Y. Khidas[2] and O. Pitois[1,*]

[1] Université Paris Est, Laboratoire Navier, UMR 8205 CNRS – École des Ponts ParisTech – IFSTTAR
cité Descartes, 2 allée Kepler, 77420 Champs-sur-Marne, France.
francois.gorlier@ifsttar.fr ; olivier.pitois@ifsttar.fr

[2] Université Paris Est, Laboratoire Navier, UMR 8205 CNRS – École des Ponts ParisTech – IFSTTAR
5 bd Descartes, 77454 Marne-la-Vallée Cedex 2, France.
yacine.khidas@u-pem.fr

* Corresponding author: Olivier Pitois, olivier.pitois@ifsttar.fr

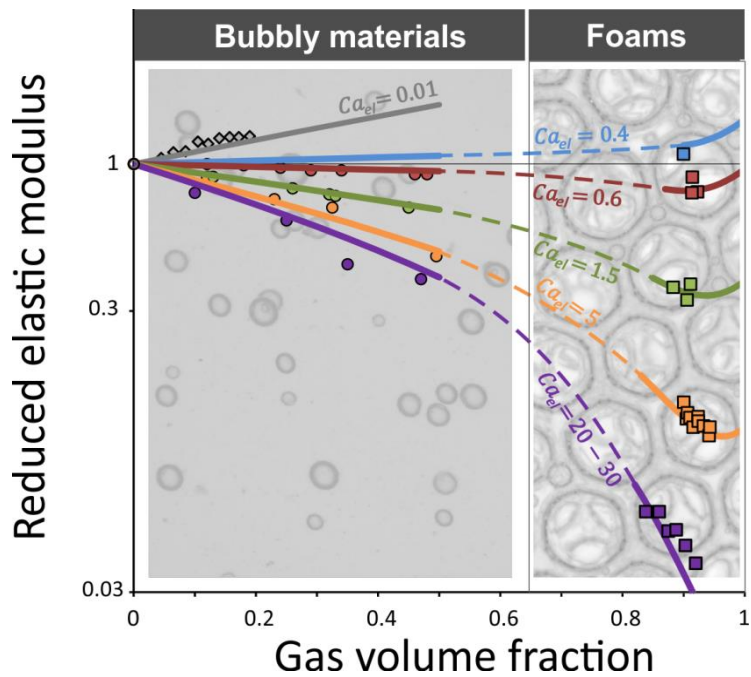

Abstract:

Elasticity of soft materials can be greatly influenced by the presence of air bubbles. Such a capillary effect is expected for a wide range of materials, from polymer gels to concentrated emulsions and colloidal suspensions. Whereas experimental results and theory exist for describing the elasto-capillary behavior of bubbly materials (i.e. with moderate gas volume fractions), foamy systems still require a dedicated study in order to increase our understanding of elasticity in aerated materials over the full range of gas volume fractions. Here we elaborate well-controlled foams with concentrated emulsion and we measure their shear elastic modulus as a function of gas fraction, bubble size and elastic modulus of the emulsion. Such complex foams possess the elastic features of both the bubble assembly and the interstitial matrix. Moreover, their elastic modulus is shown to be governed by two parameters, namely the gas volume fraction and the elasto-capillary number, defined as the ratio of the emulsion modulus with the bubble capillary pressure. We connect our results for foams with existing data for bubbly systems and we provide a general view for the effect of gas bubbles in soft elastic media. Finally, we suggest that our results could be useful for estimating the shear modulus of aqueous foams and emulsions with multimodal size distributions.

**Nomenclature:**

$\gamma$: liquid/gas surface tension

$R$: bubble radius

$E_0, G_0, G'_0$: elastic modulus of the interstitial material (emulsion)

$Ca_{e\ell} = RG'_0/\gamma$: elasto-capillary number

$\phi$: gas volume fraction in emulsion foam

$\phi_0$: gas volume fraction in the precursor foam

$\phi_c$: packing volume fraction of bubbles

$\varphi_{oil}^0$: oil volume fraction in mother concentrated emulsion

$\varphi_{oil}$: oil volume fraction in interstitial phase of emulsion foam,

$\varphi_{oil}^c$: packing volume fraction of oil droplets in oil/water emulsion

$G'(\phi, G'_0)$ or $G'(\phi)$: shear elastic modulus of emulsion foam at gas volume fraction $\phi$ (and interstitial elastic modulus $G'_0$)

$G'(\phi, 0)$: shear elastic modulus of aqueous foam at gas volume fraction $\phi$

## 1. Introduction

Aerated materials are widely encountered as products from various industries. Polymer foams, such as polystyrene or polyurethane foams are emblematic examples showing unique properties that have been proved to be useful for numerous applications. Aerated cement concrete is another example where lightness and mechanical strength have to be optimized in order to provide suitable construction materials[1]. The mechanical properties of such systems have been widely investigated in relation with the microstructure arising from the organization of bubbles and solid matrix in the material[2]. Depending on incorporated air volume fraction, the mechanical strength is estimated from the effect of holes (voids) in the bulk matrix[3,4] or from the mechanical behavior of microstructural elements forming the solid skeleton of the material[2]. It is worth noting that the effect of capillarity is not accounted for because the elastic modulus $E_0$ of the matrix, typically from 1 MPa to 10 GPa, is large compared to the surface energy effects.

It is however possible to deal with more soft elastic matrices, characterized by elastic modulus much closer to capillary pressure. This is the case for some extremely soft polymeric gels, such as polydimethylsiloxane (PDMS) and polyacrylamide (PAA)[5], as well as for several biopolymers[6,7], such as cross-linked networks of actin or fibrin, gelatin, … This is also the case for most of the so-called soft glassy materials[8], such as concentrated emulsions, dense granular suspensions and pastes. In terms of rheology, those materials exhibit a yield stress, i.e. they behave like a (visco)elastic solid until a certain critical stress is exceeded. Aerated yield stress fluids are encountered in numerous application fields, such as foods, paints, cosmetics, construction materials, … For several of those materials, capillary effects has been evidenced. Style et al.[9] showed how capillary inclusions (here small liquid drops) can either stiffen or soften a polymer matrix. Ducloué et al.[10] demonstrated that the shear elastic modulus of concentrated emulsions can be tuned by incorporating well-chosen air bubbles. Such studies were mostly focused on 'bubbly' systems, for which the volume fraction $\phi$ of capillary inclusions is smaller than $\phi_c \approx 0.64$, the packing volume fraction of spherical inclusions. However, systems characterized by $\phi > \phi_c$, are also widely encountered. For example, biomedical foams are used for tissue engineering applications[11], and food industry develops more and more foamed products, from alginate and gelatin foams to whipped creams[12]. Foamed cement[1], geopolymer[13] or plaster[14] fresh pastes are also examples for foamy soft materials. How matrix elasticity couples with bubble elasticity in such materials? Whereas foams have been studied mainly in situations where either the solid skeleton or the bubble elasticity was governing solely their mechanical behavior, i.e., $E_0 \gg 2\gamma/R$ or $E_0 = 0$ respectively (where $2\gamma/R$ is the bubble capillary pressure), there is no available experimental or theoretical result for predicting this behavior as both contributions count, i.e., for $E_0 \approx 2\gamma/R$. The latter is precisely the situation we investigate in this paper. We elaborate well-controlled foams made with concentrated emulsion and we measure their shear elastic modulus as a function of gas volume fraction, bubble size and elastic modulus of the interstitial emulsion. We show that emulsion foams possess the elastic features of both the matrix and the bubble assembly, and that the elastic modulus is governed by two parameters: the gas volume fraction and the elasto-capillary number.

## 2. Materials and methods

Emulsion foams are prepared by mixing aqueous foam and emulsion following the procedure sketched in figure 1. The first step of the preparation is the production of precursor aqueous foam with well-controlled bubble size and gas volume fraction (see figure 1a). Foaming liquid (distilled water 70% w/w, glycerol 30% w/w, and tetradecyltrimethylammonium bromide at a concentration 5 g.L$^{-1}$) and perfluorohexane-saturated nitrogen are pushed through a T-junction allowing controlling the bubble size by adjusting the flow rate of each fluid. Several bubble sizes were obtained: $R = 150$, 270, 300 and 400 µm, with polydispersity $\Delta R/R \approx \pm 5\%$. Produced bubbles are collected in a glass column and gas fraction is set to an approximately constant value $\phi_0$ over the foam column by imbibition from the top with foaming solution (see figure A1). As shown in the following, $\phi_0$ has to be known as accurately as possible. From careful work on the reproducibility of foaming and drainage conditions in the column, we were able to define a target value within a range $\phi_0 \pm \Delta\phi_0$ with $\Delta\phi_0 = 0.001$ and $\phi_0$ set to a value ranging from 0.985 to 0.995 (see appendix).

Secondarily, we use a concentrated oil-in-water mother emulsion which has been produced in advance using a Couette emulsifier. The oil is a silicon oil (V350, Chimie Plus) and the water solution is composed of distilled water 50% w/w, glycerol 50% w/w, and tetradecyltrimethylammonium bromide at a concentration 30 g.L$^{-1}$). Note that continuous phases of the foam and the emulsion are almost the same. The oil volume fraction is $\varphi_{oil}^0$ = 0.85 and the radius of the oil droplets measured by laser granulometry is around 1 to 2 µm (the polydispersity is around 20%). Note also that the oil/water/surfactant system chosen for this study is known to produce very stable mixtures of foam and emulsion[10,15–17].

In the next step, precursor foam and emulsion are mixed in a continuous process thanks to a mixing device based on flow-focusing method (see figure 1b). The mixing device allows tuning the flow rates of both the foam ($q_{pf}^*$) and the emulsion ($q_0$), as well as introducing additional foaming solution at a flow rate $q_s$ in order to dilute the mother emulsion. Note that the efficiency of the dilution process has been checked separately by measuring the elastic modulus $G_0'$ of resulting diluted emulsions as a function of oil volume fraction. In fact we prepared three mother emulsions, and for each one the curve $G'_0(\varphi_{oil})$ was determined. Typical example is plotted in figure 2a, showing that $G'_0(\varphi_{oil})$ curves can be described by the relationship $G'_0 = g\varphi_{oil}(\varphi_{oil} - \varphi^c_{oil})$, where $g$ and $\varphi^c_{oil}$ are close to 5100 Pa and 0.63 respectively, which is fully consistent with previous results[18]. The device produces emulsion foams with target values for gas volume fraction $\phi^*$ and oil volume fraction $\varphi^*_{oil}$. Note also that bubble size is conserved during the mixing step. According to parameters defined above, target values are $\phi^* = q^*_{pf}\phi_0/(q^*_{pf} + q_0 + q_s)$ and $G'^*_0 = g\varphi^*_{oil}(\varphi^*_{oil} - \varphi^c_{oil})$ with $\varphi^*_{oil} = q_0\varphi^0_{oil}/(q_0 + (1-\phi_0)q^*_{pf} + q_s)$. Note that $\varphi^*_{oil}$ is the oil volume fraction in the interstitial volume of the foam (between bubbles), and the oil volume fraction in the whole foam volume is given by $\phi^*_{oil} = \varphi^*_{oil}(1-\phi) = q_0\varphi^0_{oil}/(q_0 + q^*_{pf} + q_s)$. During the dilution/mixing step, liquid flow rates $q_0$ and $q_s$ are imposed, but the target flow rate $q^*_{pf}$ is subjected to gas compressibility effects. In order to evaluate and to account for such induced deviations with respect to target values, we measured the actual value of the gas volume fraction in emulsion foams, $\phi = 1 - \rho/\langle\rho_{em}\rangle$, where $\rho$ is the density of the sample deduced from weight measurements, and $\langle\rho_{em}\rangle = 1.005$ is the median value of density for the mixed emulsions (densities of oil and water/glycerol solutions are 0.97 and 1.13 (for 50/50) or 1.08 (for 70/30) respectively). From $\phi$ we determine the actual value $q_{pf} = (q_0 + q_s)\phi/(\phi_0 - \phi)$ as well as the resulting oil volume fraction $\varphi_{oil} = q_0\varphi^0_{oil}/(q_0 + (1-\phi_0)q_{pf} + q_s)$. Following such a method induced the following maximum relative errors: $\Delta\varphi_{oil}/\varphi_{oil} \approx \pm 3\%$, $\Delta G'_0/G'_0 \approx \Delta G'/G' \approx 25\%$ and $\Delta\phi/\phi \approx \pm 1\%$.

A typical example of emulsion foams we prepared is presented in figure 1c. Note that emulsion is contained in the Plateau border network, not in the thin foam films (figure 1d). The efficiency of the preparation procedure is demonstrated in figure 2b, where reachable parameters $G'_0$ – the modulus of the interstitial emulsion – and $\phi$ – the gas volume fraction – are plotted, as well as actuals parameters for samples prepared for this study.

Resulting emulsion foams are continuously poured into the measurement cell (cup geometry: height = 7 cm and diameter = 37mm). After this filling step, a six-bladed vane tool (height = 6 cm and diameter = 25mm) is inserted into the foam cell. The rheology measurement procedure starts with a stress-controlled rheometer (Malvern kinexus ultra+): elastic and loss moduli are measured as a function of increased shear strain amplitude $\epsilon$ at a fixed frequency of 1Hz. Note that (1) emulsion foams were not observed to drain, (2) to avoid slippage on the cell wall as the shear stress is applied, the cell surface has been striated to jam the bubbles; (3) depending on the bubble size, the minimal gap in the vane-cup geometry represents from 8 to 20 bubble diameters; (4) the presence of perfluorohexane inside bubbles strongly reduces the foam ripening rate[19] which allows aging affects to be ignored over the time scale of measurements.

## 3. Results

Before focusing on elastic modulus of emulsion foams, we present the comparative typical rheological behaviors of foams, emulsions and emulsion foams (see figure 3). Those three materials own to the same class of materials, exhibiting a linear elastic regime for strain amplitudes smaller than a critical value $\epsilon_{crit} \approx 0.1$, from which the shear elastic modulus decreases strongly – and the loss modulus increases simultaneously –, accounting for the onset of flow in the system. In terms of rheological behavior, emulsion foams can be compared directly to their constituents, without any particular or anomalous effect, which turns them into ideal systems for studying the contribution of the interstitial phase on foam elasticity. In the following, the elastic modulus $G'$ we consider corresponds to the value obtained at small strain amplitudes, i.e. $\epsilon \lesssim 0.001$.

The interest of the elaboration method is presented in figure 4, where $G'$ is plotted as a function of the bubble radius $R$ (figure 4a), of the gas volume fraction (figure 4b) and the modulus of the interstitial emulsion $G'_0$ (figure 4c). Note that in each case, other control parameters are fixed. The elastic modulus of emulsion foams is shown to decrease with $R$, which is the hallmark of aqueous foams. However, increasing the gas volume fraction results in decreasing $G'$. This effect is contrary to the evolution observed for classical aqueous foams and it can be, therefore, attributed to the interstitial emulsion. Figure 4c also shows another effect of the emulsion: increasing $G'_0$ makes $G'$ increase as well, but to a lesser extent. The whole set of data is now presented in figure 5, where the reduced modulus $G'(\phi, G'_0)/G'(\phi, 0)$ is plotted as a function of $\phi$ (figure 5a) and $G'_0$ (figure 5b) for investigated bubble sizes. $G'(\phi, 0)$ is the elastic modulus of the corresponding aqueous foam, i.e. the emulsion-free foam with same bubble size and same gas volume fraction. The latter can be estimated by the classical relationship[20]: $G'(\phi, 0) = a\gamma\phi(\phi - \phi_c)/R$ where $\gamma \approx 0.035$ mN/m is the bubble surface tension in the water/glycerol continuous phase, and $\phi_c = 0.64$ is the packing volume fraction of spherical bubbles. As we measured $G' \approx 60$ Pa for the drained emulsion-free foam (with $R$ = 300 μm), which is characterized by an average gas volume fraction equal to $\phi_{eq} \approx 0.97$, chosen parameter is $a \approx 1.6$, which falls within the range of accepted values[21]. It is shown in figure 5 that due to the interstitial complex material, the shear modulus of emulsions foams can be as high as twice the modulus of the corresponding aqueous foam. The rescaling of $G'$ with $G'(\phi, 0)$ is not sufficient to obtain a single curve as a function of $\phi$ or $G'_0$ for all the bubble sizes, suggesting that the contributions of both the bubble assembly and the emulsion have to be accounted for.

## 4. Discussion

In the following we propose physical elements for describing the elastic modulus of emulsion foams. As presented in the previous section, contributions of both the bubble assembly and the interstitial concentrated emulsion have to be considered. Basically, the contribution of contacting bubbles can be estimated from the shear elastic modulus $G'_b$ of the corresponding aqueous foam, i.e. with the same gas volume fraction[20]:

$$G'_b = G'(\phi, 0) = a \frac{\gamma}{R} \phi(\phi - \phi_c) \quad (1)$$

where $a \approx 1.6$ and $\phi_c \approx 0.64$. The space between bubbles is filled with concentrated emulsion which is known to behave as a solid for small strains. One can estimate the corresponding elastic contribution from results for mechanics of solid cellular materials, as reviewed by Gibson and Ashby[2]:

$$G'_{sk} = b G'_0 (1 - \phi)^n \quad (2)$$

where, for open cell foams (the emulsion is not present in the thin aqueous films), the parameters $b$ and $n$ have been shown to be close to 1 and 2 respectively. In equation 2, $G'_0$ represents the bulk shear modulus of the solid matrix forming the skeleton of the open cell foam. Recent experiments, performed on foams made with granular matter as interstitial material, have suggested that the emulsion can be considered as a continuous matrix filling foam's Plateau borders and nodes[22]. Both contributions can be combined in order to give an estimate of the global elastic modulus:

$$G'_{th} = G'_b + G'_{sk} + \psi \quad (3)$$

Note that in equation 3 we have introduced a parameter $\psi \geq 0$ accounting for coupling effects arising from mechanical interactions between the emulsion matrix and bubbles. Such an interaction could be related to differences in the local deformation fields of both constituents. One can also consider that the presence of the interstitial yield stress material alters the equilibrium configuration of the bubbles with respect to the corresponding aqueous foam.

Experimental data are now compared to equation 3 with $\psi = 0$. Note that taking $\psi = 0$ involves considering that contributions of bubbles and emulsion superimpose ideally. Such a situation is presented in figure 6a, showing how interstitial emulsion reinforces the shear modulus of foams, especially for gas volume fractions within the range 0.8-0.9. In figure 6b, $G'$ is plotted as a function of $G'_{th}$, showing a poor global agreement as well as significant deviations for a large number of investigated systems. This highlights that the coupling term has to be considered. As discussed above, $\psi$ is expected to depend on interactions between bubbles and interstitial emulsion, suggesting the following involved parameters: $G'_0$, $\gamma$, $R$ and $\phi$. We introduce the elasto-capillary number $Ca_{e\ell} = RG'_0/\gamma$ and we expect the reduced coupling term, i.e. $\psi/G'_0$ or $\psi/(\gamma/R)$, to depend on both $Ca_{e\ell}$ and $\phi$. Note that for $Ca_{e\ell} \gg 1$, we expect $G'_{sk} \gg G'_b$ and $G'_{th} \approx G'_{sk}$, which means that $\psi \approx 0$. Moreover, as $\phi \to 1$ we expect $G'_{sk} \ll G'_b$ and $G'_{th} \approx G'_b$, which also means that $\psi \approx 0$. The modelling of the coupling term is a huge theoretical task, as suggested by theoretical work devoted to bubbly systems[23], and it is far beyond the scope of the present paper. Now we try to show that collected data can be described with a function of $Ca_{e\ell}$ and $\phi$. Therefore, we seek for a simple analytical form complying with the expected behavior for $Ca_{e\ell} \gg 1$ and $\phi \to 1$. We propose arbitrarily the following form:

$$\psi/G'_0 = 15(1-\phi)^2(2\phi-1)Ca_{e\ell}^{-2/3} \quad (4a)$$

$$\psi/(\gamma/R) = 15(1-\phi)^2(2\phi-1)Ca_{e\ell}^{1/3} \quad (4b)$$

Agreement of equations 3-4 with experimental data can be considered from figure 7a, where $G'$ is plotted as a function of $G'_{th}$, as well as from figure 4. Note that (i) observed deviations are smaller than 15% and the average deviation is equal to 5%, (ii) equation 4 should not be used outside of the ranges of investigated parameters, i.e. $Ca_{e\ell} \lesssim 0.5$ and $\phi \lesssim 0.85$. Equations 3-4 are now used to provide a clear picture for foam strengthening effect due to interstitial elasticity. $G'(\phi, G'_0)/G'(\phi, 0)$ is plotted in figure 7b as a function of the elasto-capillary number for several values of the gas volume fraction. It is shown that whereas foams with 95% gas are only weakly influenced by interstitial elasticity, elastic strengthening is significant for foams with 90% and 85% of gas.

Literature gives several examples of studies focused on the evolution of elastic modulus of soft elastic media containing bubble or drop inclusions[10,16,17,24,25]. How results we present for foams compare with elasticity of bubbly materials? For such aerated media, the parameter of interest is the reduced elastic modulus $G(\phi)/G_0$, which accounts for the effect of capillary inclusions in the elastic matrix. To this regard, it has been shown that within the range $0 \le \phi \lesssim 0.5$, $G(\phi)/G_0$ depends on the elasto-capillary number $Ca_{e\ell}$ through the equation[23]:

$$\frac{G(\phi)}{G_0} = 1 - \frac{\phi(2Ca_{e\ell}-1)}{1+6Ca_{e\ell}/5+2\phi(2Ca_{e\ell}-1)/5} \quad (5)$$

Note that for $Ca_{e\ell} = 0.5$, the effective elasticity of bubble inclusions is the same as the embedding matrix and $G(\phi)/G_0 = 1$, whatever the volume fraction of inclusions. In figure 8 we plot $G(\phi)/G_0$ as a function of $\phi$ for several $Ca_{e\ell}$ values. We report available data for bubbly materials, i.e. Ducloué et al.[10] and Style et al.[9], and for soft elastic foams, i.e. Khidas et al.[25], as well as our data. We plot also the theoretical prediction for bubbly materials, i.e. equation 5, as well as equations 3-4. It is manifest that results for soft solid foams are in line with results from bubbly materials, i.e. they follow the same global trend imposed by the elasto-capillary number at small gas volume fractions, but amplified for large $\phi$ values. In fact, the elasticity of the contacting bubble assembly becomes apparent through the presence of a local minimum, the latter occurring for a $\phi$ value that increases as $Ca_{e\ell}$ increases. Note that for a given value $\phi$, the transition from strengthening (i.e. $G(\phi)/G_0 > 1$) to softening (i.e. $G(\phi)/G_0 < 1$) occurs for a $Ca_{e\ell}$ value close to 0.5. For foams, this behavior is correctly described by equations 3-4. Note also that data from Khidas et al.[25] obtained for gelatin foams are well-described by equations 3-4, so that the latter provide a good estimate of foam modulus over almost two orders of magnitude for $Ca_{e\ell}$ values.

## 5. Conclusion

Solid foams and simple liquid foams have been widely studied[2,20] and their mechanical behavior is now quite well understood. In contrast, understanding the behavior of soft solid foams, for which mechanics is expected to be related to both capillary contribution of the bubbles and solid elasticity of the interstitial skeleton, is more challenging. We have proposed a new approach for studying those complex systems: we have elaborated dedicated emulsion foams with unequaled control of all parameters, i.e. bubble size, gas volume fraction, and elastic modulus of the interstitial emulsion.

The global elastic modulus measured in the linear regime for such a foamy system has been shown to be determined by a complex interplay of the control parameters, i.e. the elastic modulus of the interstitial material, the bubble size and the gas volume fraction. We have shown, however, that such a behavior can be described using a simple analytical form of the elasto-capillary number $Ca_{e\ell}$ and the gas volume fraction $\phi$, over the whole range of investigated parameters, i.e $0.85 \lesssim \phi \lesssim 0.95$ and $0.4 \lesssim Ca_{e\ell} \lesssim 10$.

Although based on emulsion foams, the present research is based on the use of the elasto-capillary number as a control parameter, which allows for our results to be applied to numerous foamy systems. In this regard, eqs 3-4 represent an interesting tool for practitioners, allowing for the elastic modulus of any foamed system to be predicted, as soon as the elasto-capillary number is known. Such a predictive tool can be useful for the development aerated soft elastic materials, such as those encountered in tissue engineering applications[11] and food industry[12], for example. For example, elastic properties of soft solid foams can be tuned accurately by varying equivalently either the bubble size, or the modulus of the soft solid, or the gas fraction. The potential of such a general approach is demonstrated in figure 8, where our results have been connected with previous works focused gelatin foams[25] and soft solid media containing bubble or drop inclusions[10,16,17,24]. Moreover, we have shown that the behavior of investigated foams is in line with results for bubbly solids: increasing gas volume fractions amplifies the effect of the elasto-capillary number observed at small gas volume fractions. This behavior stops for highest gas volume fractions, where collective bubble elasticity dominates finally. It is worth noting that covering the intermediate range of gas volume fractions including the packing volume fraction of spherical bubbles, i.e. marking the transition between bubbly and foamy materials, would be useful for providing better understanding of elasticity in aerated soft materials. In the regard, figure 8 is appealing for dedicated theoretical work.

Finally, we suggest that results presented in figure 8 could be used for estimating the effect of polydispersity on shear modulus of foams and emulsions. This can be exemplified by the case of bidisperse emulsions, as studied experimentally by Foudazi et al.[26]. In such a case, the emulsion can be considered as mixture of large droplets (radius $R_L$ and volume fraction $\phi_L$) and small droplets (radius $R_S$ and volume fraction $\phi_S$). Using equation 5 one can estimate the evolution of $G_0$, the elastic modulus of small droplets emulsion, as a function of the elasto-capillary number $Ca_{e\ell} = G_0/(\gamma/R_L)$ and the volume fraction of large droplets, i.e., $\phi \equiv \phi_L$. Inset in figure 8 shows that such an approach makes sense although it seems difficult to discuss more the results as "large" and "small" droplets were intrinsically polydisperse with partial overlapping. This point would require a dedicated study.

**Appendix**

As explained in the experimental part, the success of our generation method is based on appropriate control of liquid content in the precursor foam. We address this issue through both numerical simulations of the foam production process and dedicated experiments.

We start the numerical approach by considering Darcy's law applied to aqueous foam[20]:

$$\vec{v} = \frac{k(\phi)}{\eta}\left(\rho\vec{g} + \vec{\nabla}\Pi(\phi)\right) \quad \text{(eq. A1)}$$

where $v$ is the superficial liquid velocity, i.e. liquid flow rate divided by the cross-section area of the foam column $S$ = 5 cm², $k(\phi)$ and $\Pi(\phi)$ are respectively the foam permeability and the foam osmotic pressure. Using the continuity equation, $-\partial\phi/\partial t + \vec{\nabla}\cdot\vec{v} = 0$, one obtains the so-called drainage equation:

$$-\frac{\partial \phi}{\partial t} + \vec{\nabla} \cdot \left[\frac{k(\phi)}{\eta}\left(\rho\vec{g} + \vec{\nabla}\Pi(\phi)\right)\right] = 0 \quad \text{(eq. A2)}$$

Expressions for $k(\phi)$ and $\Pi(\phi)$ are given by[27,28]:

$$k(\phi) \simeq \frac{R^2 \phi^{3/2}}{425(1-2.7\phi+2.2\phi^2)^2} \quad \text{(eq. A3)}$$

$$\Pi(\phi) \simeq 7.3 \frac{\gamma}{R} \frac{(0.74-\phi)^2}{\phi^{1/2}} \quad \text{(eq. A4)}$$

Equation A2 is solved by the finite difference method (FDM). The following boundary conditions are used: $\phi(0)$ = 0.74 (monodisperse ordered – CFC – foam) and $d\Pi/dz$ = 0 at $z$ = 0 (bottom of the column), $v_i = Q_i/S$ is the liquid velocity imposed at the top of the foam for simulating the imbibition process.

The foam production is composed of three parts. (1) Generation step. A foam volume $V_f$, or equivalently a foam height $h_f = V_f/S$, is generated at a gas flow rate $Q_g$: a foam slice of thickness $\Delta z \simeq 10R$, i.e. $\Delta z \simeq h_f/100$, is inserted at foam bottom at each time interval $\Delta t = \Delta z S/Q_g$, i.e. existing foam slices are shifted upward from a distance equal to $\Delta z$. Gas volume fraction of the inserted foam slice is $\phi(0)$ = 0.74. During $\Delta t$, foam is draining, involving a net liquid flux for each foam slice at height $z$ and a resulting new value $\phi(z)$ calculated according to equation A2 (see figure A1). (2) Drainage step. The generation process is stopped and the foam column is let to drain for a duration $\Delta t_i$ with liquid imbibition from the top at flow rate $Q_i$ (see figure A1). (3) Foam delivery step. Foam flushing out of the column at a flow rate $Q_d$ is simulated by "removing" a foam slice $\Delta z$ from the top at each time interval $\Delta t = \Delta z S/Q_d$. Liquid volume fraction of the removed foam slice is measured. During $\Delta t$, each foam slice at height $z$ is draining according to equation A2. Note that the imbibition flow rate $Q_i$ can be maintained at the same value as in the previous step or set to zero.

In order to validate the numerical tool, we performed specific measurements for the liquid content of precursor foams delivered from our setup. Note that the delivery conditions of precursor foam when mixed with emulsion during the emulsion foam production are 60 cm³ within 20 min. Such conditions cannot be applied for measuring accurately the volume of the emulsion-free precursor foam due to the absence of the emulsion stabilizing effect against osmotic ripening – due to perfluorohexane – and partial collapse. Therefore, those measurements are performed at high flow rate for foam delivery (60 cm³ within 4 min). Simulations obtained with three sets of parameters are compared with experimental results in figure A2. Within experimental error for bubble size, good agreement is obtained for the three testing configurations. This suggests that numerical simulations can be used without any modification for optimizing the generation process within conditions for emulsion foams production, i.e. 60 cm$^3$ within 20 min. Figure A3 presents such results for two investigated bubble sizes, showing that precursor foams exhibit approximately constant gas volume fractions $\phi_0$, i.e. $\Delta\phi_0 \lesssim 0.001$.

**Acknowledgements**

We thank D. Hautemayou and C. Mézière for technical support. We gratefully acknowledge financial support from Agence Nationale de la Recherche (Grant no. ANR-13-RMNP-0003-01) and French Space Agency (convention CNES/70980). We thank X. Chateau and G. Ovarlez for fruitful discussions.

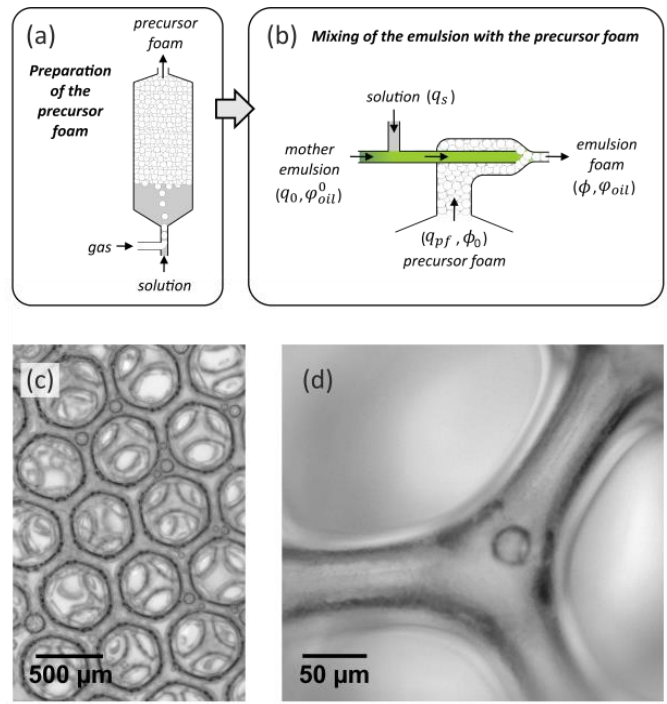

Fig. 1: Elaboration method for emulsion foams. (a) Precursor aqueous foam is generated and stabilized in a glass column. (b) Then the foam is pushed toward a device where it is mixed with a concentrated emulsion. Note that the mixed emulsion results from the dilution of a mother emulsion with (foaming) solution. Controlling the entry flow rates allows tuning the volume fractions of the constituents in the produced emulsion foam. (c) Image of emulsion foam as viewed with a microscope. (d) Magnification on the structure of emulsion foams showing the 'granular' aspect of Plateau borders due to interstitial emulsion. Note that foam films are transparent and do not contain emulsion droplets.

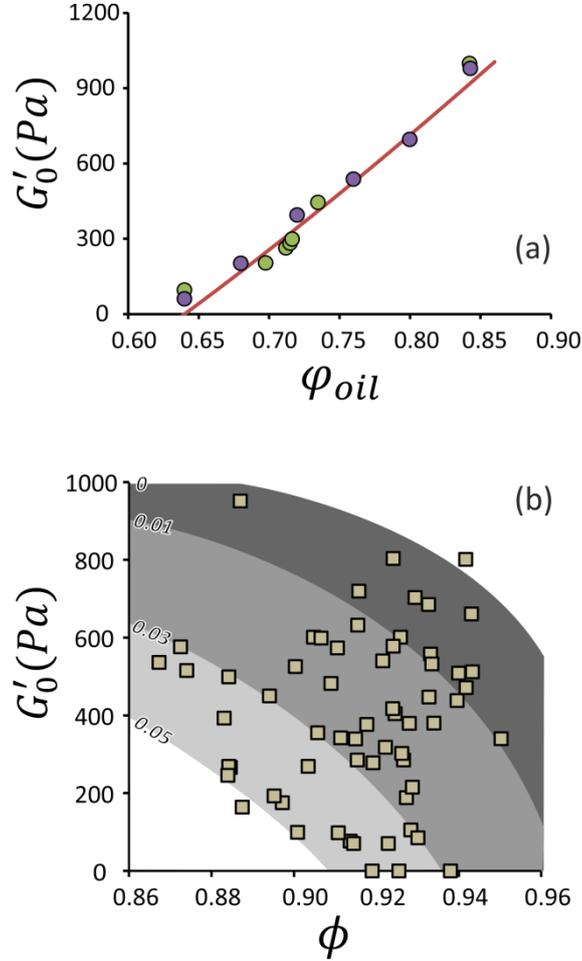

Fig. 2: (a) Oil-in-water emulsions we used are characterized by a clear relation between shear elastic modulus $G'_0$ and oil volume fraction $\varphi_{oil}$. Such a relation, which can be described by $G'_0 = g\varphi_{oil}(\varphi_{oil} - \varphi^c_{oil})$, is used to control the shear modulus of the interstitial emulsion through tuning of its oil concentration. Typical values for $g$ and $\varphi^c_{oil}$ are 5100 Pa and 0.63 respectively. (b) Thanks to the elaboration method (see figure 1), parameters $G'_0$ and $\phi$ – the gas volume fraction – can be tuned as shown by symbols representative of investigated samples. Shaded areas correspond to reachable ranges for those parameters, as defined by (see figure 1 for details about notations): $\phi = q_{pf}\phi_0/(q_{pf} + q_0 + q_s)$ and $G'_0 = g\varphi_{oil}(\varphi_{oil} - \varphi^c_{oil})$, where $\varphi_{oil} = q_0\varphi^0_{oil}/(q_0 + (1 - \phi_0)q_{pf} + q_s)$. These relations have been plotted with parameters $\phi_0 = 0.995$ and $q_s = 0; 0.01; 0.03$ and $0.05$, as indicated in the figure.

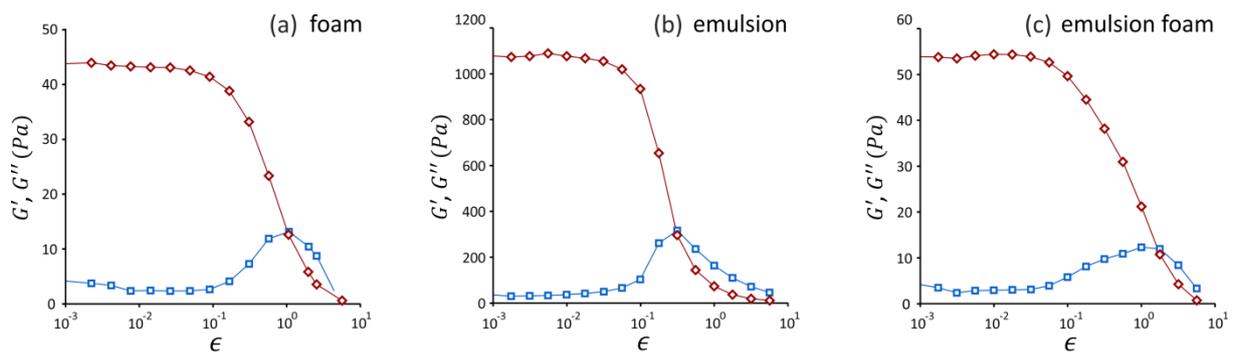

Fig. 3: Typical examples of results for shear elastic (storage) modulus $G'(\epsilon)$ (diamonds) and viscous (loss) modulus $G''(\epsilon)$ (squares) as a function of strain amplitude of aqueous foam (a), emulsion (b) and emulsion foam (c).

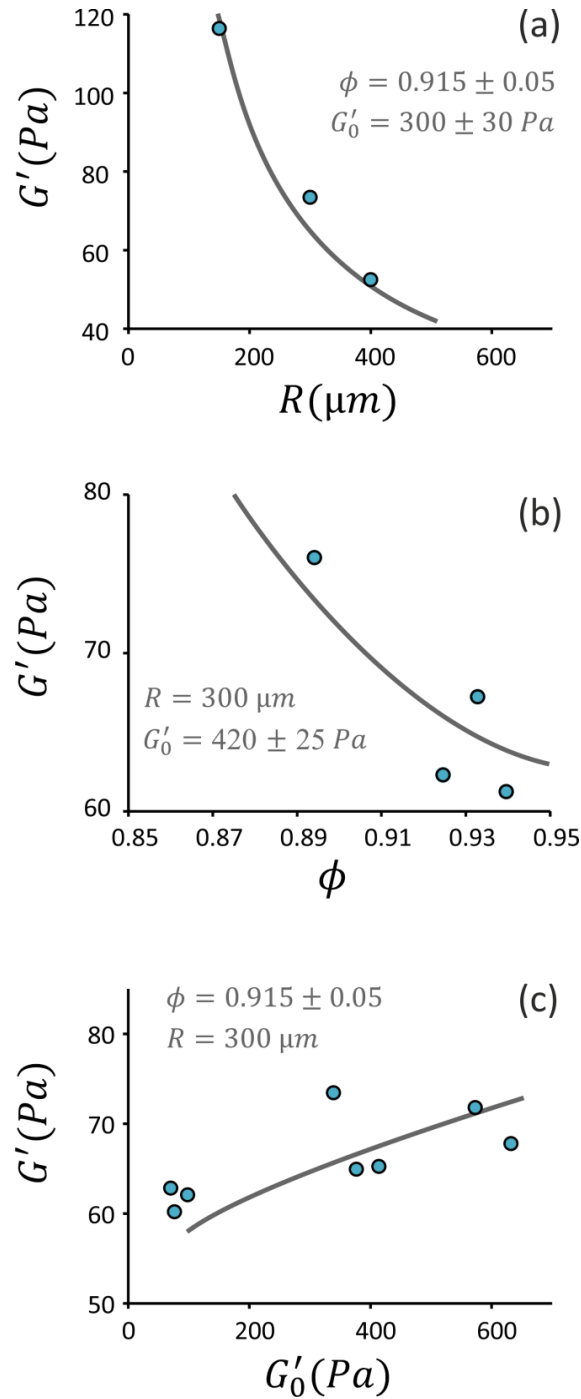

Fig. 4: Shear elastic modulus of emulsion foams as a function of control parameters. (a) Bubble radius $R$, as gas fraction and shear elastic modulus of the interstitial emulsion are fixed. (b) Gas volume fraction $\phi$, as bubble size and shear elastic modulus of the interstitial emulsion are fixed. (c) Shear elastic modulus of the interstitial emulsion $G'_0$, as bubble size and gas volume fraction are fixed. In plots a-c, continuous lines represent equations 3-4.

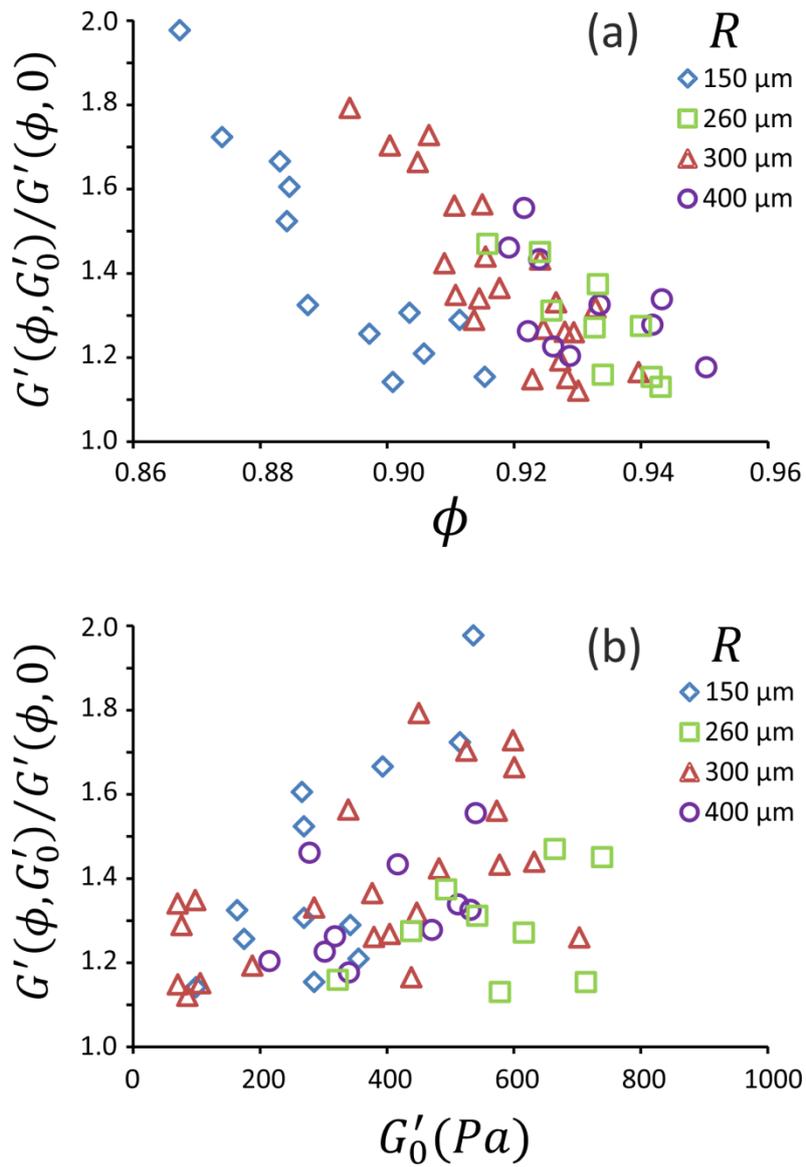

Fig. 5: Shear elastic modulus of emulsion foams divided by the shear elastic modulus of the corresponding emulsion-free aqueous foam (with the same bubble size and the same gas volume fraction) (a) as a function of the gas volume fraction $\phi$ for several bubble size values; (b) as a function of the shear elastic modulus $G'_0$ of the interstitial emulsion, for several values of the bubble size.

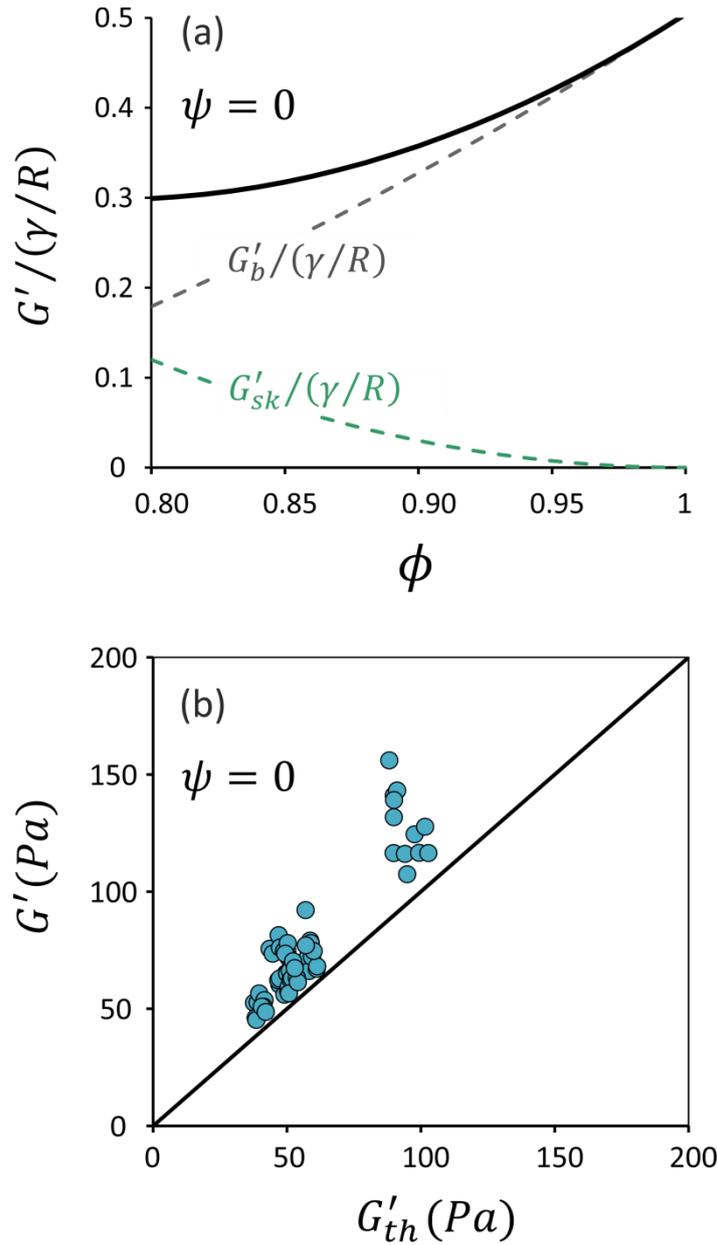

Fig. 6: (a) Shear elastic modulus of emulsion foams (continuous line) as predicted by equation 3, using $\psi = 0$. This situation corresponds to ideal superimposition of contributions of both bubbles and interstitial emulsion elasticities (dashed lines). Note that bubble elasticity increases as gas volume fraction increases (see equation 1) whereas elasticity induced by the confined emulsion decreases in the same time. (b) Comparison of experimental data for the shear elastic modulus $G'$ of emulsion foams with values $G'_{th}$ predicted by equation 3, using $\psi = 0$. The continuous line represents the equality $G' = G'_{th}$.

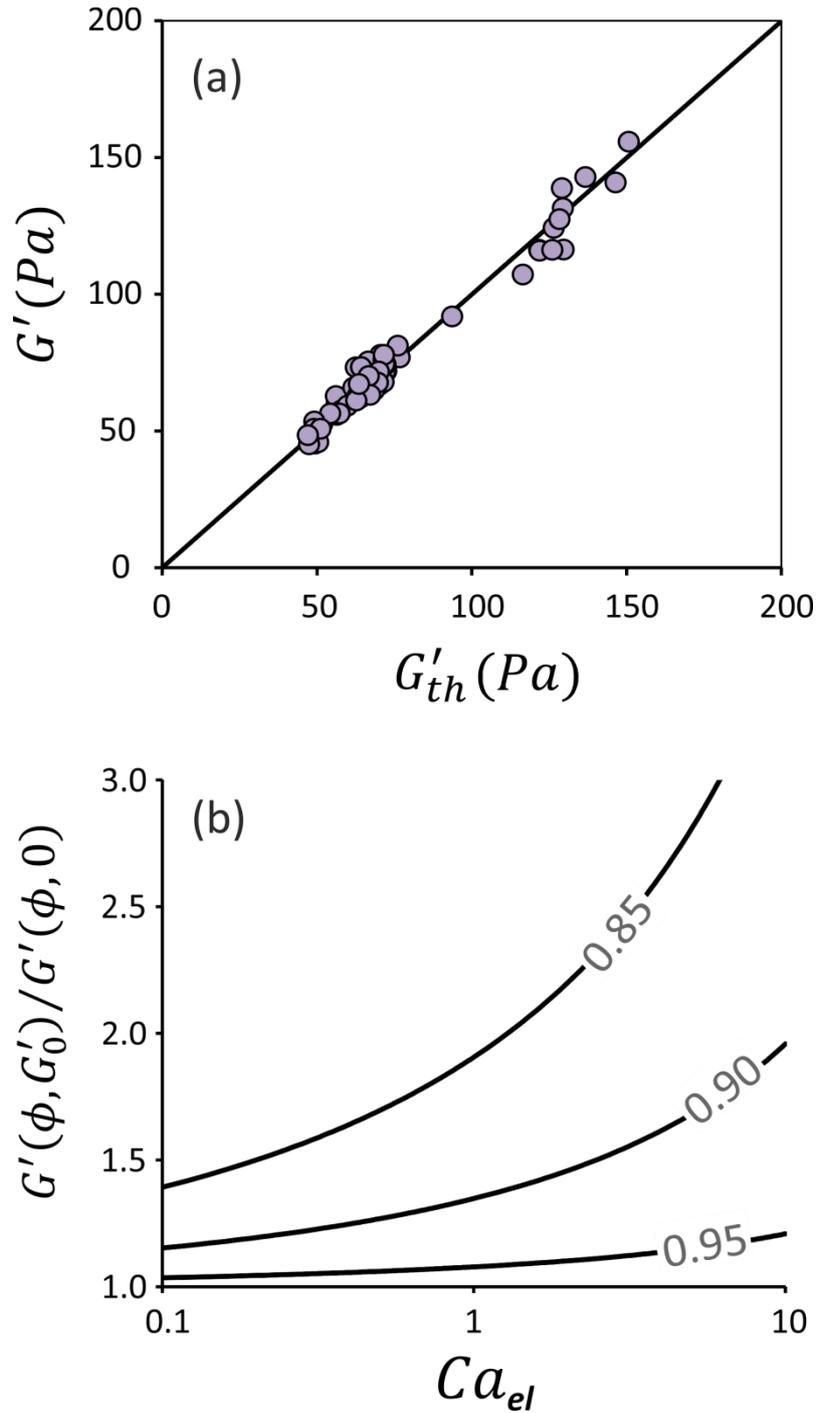

Fig. 7: (a) Comparison of experimental data for the shear elastic modulus $G'$ of emulsion foams with values $G'_{th}$ predicted by equations 3-4. The continuous line represents the equality $G' = G'_{th}$. (b) Shear elastic modulus of emulsion foams divided by the shear elastic modulus of the corresponding emulsion-free aqueous foam (with the same bubble size and the same gas volume fraction) as predicted by equations 3-4 as a function of the elasto-capillary number $Ca_{e\ell}$, for several values of the gas volume fraction $\phi$.

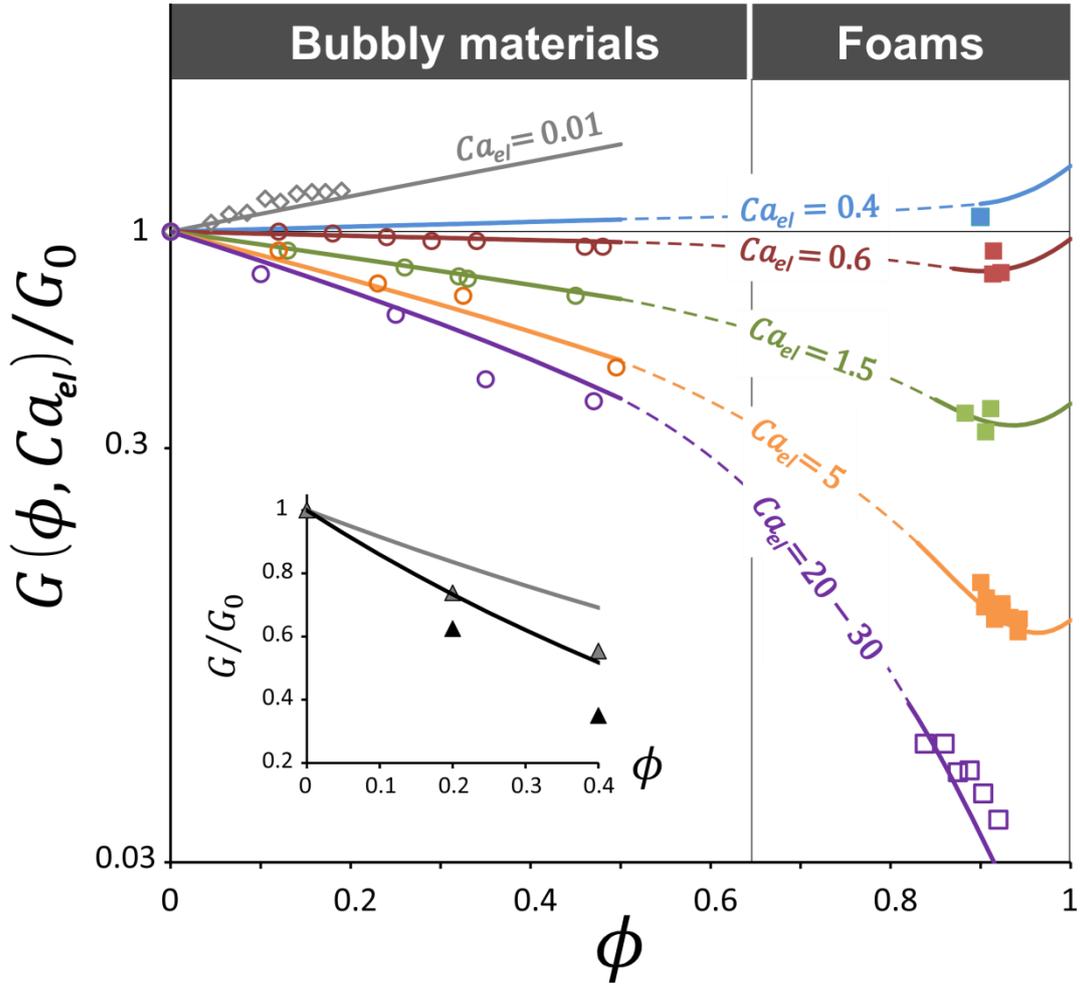

Fig. 8: Shear elastic modulus of bubbly and foamy solids divided by the elastic modulus of the bulk material as a function of gas volume fraction $\phi$, for several values of the elasto-capillary number $Ca_{e\ell}$. Filled squares: this work; empty squares: data from Khidas et al.[25]; circles: data from Ducloué et al.[10]; diamonds: data from Style et al.[9]; continuous lines: equation 5 for $\phi < 0.5$ and equations 3-4 for $\phi > 0.85$; dashed lines are guides for the eye. Inset: Data adapted from Foudazi et al.[26] showing the shear modulus of bidisperse emulsions, i.e. mixture of "small" and "large" droplets, as a function of the volume fraction of "large" droplets. Note that in such a case, $G_0$ is the shear modulus for the emulsion made with the "small" droplets. The estimated corresponding $Ca_{e\ell}$ values are equal to 2 (grey triangles) and 10 (black triangles). Continuous lines represent equation 5 for those two $Ca_{e\ell}$ values.

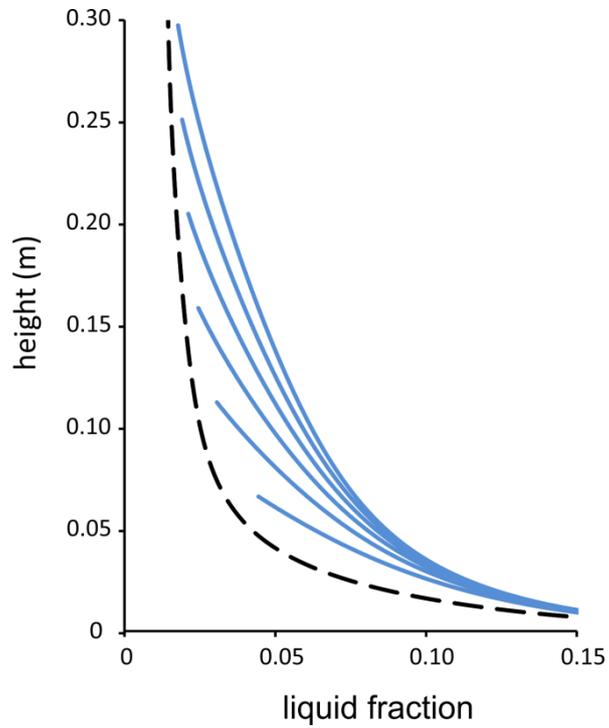

Fig. A1: Example of the results obtained with the numerical approach. Continuous lines show the evolution of the vertical profile of the liquid volume fraction for several times: (from bottom to top) t = 9, 15, 22, 28, 36 and 40 min. The latter time corresponds to the end of the generation process. Parameters are the following: $R$ = 140 µm, $Q_g$ = 3.5 cm³/min, $Q_i$ = 0.6 cm³/h. The dashed line shows the vertical profile at time equal to 100 min, i.e. 1 hour after the end of the foam generation process with $Q_i$ = 0.6 cm³/h.

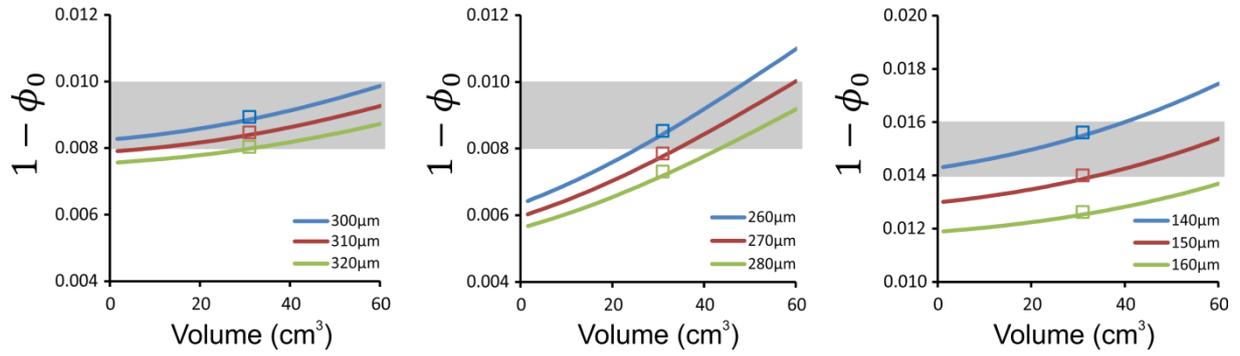

Fig. A2: Liquid volume fraction of precursor foams as delivered from the generation column within testing conditions. Shaded areas correspond to the global (averaged) liquid fraction measured over 60 cm³ delivered by the generation column. Solid curves correspond to numerical simulations. They give the foam liquid volume fraction at the column outlet as a function of the delivered foam volume. Empty squares show the average values of liquid volume fraction over the full profile. The foam production parameters are the following: (a) $Q_g$ = 7 cm³/min, $Q_i$ = 1.2 cm³/h, $\Delta t_i$ = 5 min, $Q_f$ = 15 cm³/min. (b) $Q_g$ = 7 cm³/min, $Q_i$ = 0.5 cm³/h, $\Delta t_i$ = 5 min, $Q_f$ = 15 cm³/min. (c) $Q_g$ = 3.5 cm³/min, $Q_i$ = 0.6 cm³/h, $\Delta t_i$ = 60 min, $Q_f$ = 15 cm³/min.

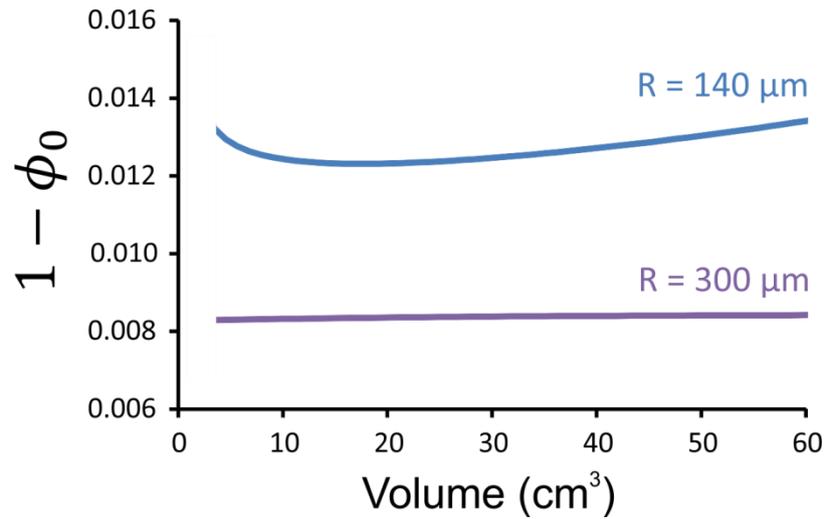

Fig. A3: Liquid volume fraction of precursor foams as delivered from the generation column within experimental conditions for the production of emulsion foams. Parameters are the following: (1) $R$ = 140 µm, $Q_g$ = 3.5 cm³/min, $Q_i$ = 0.6 cm³/h during foam generation, $\Delta t_i$ = 60 min, $Q_f$ = 3.5 cm³/min, $Q_i$ = 0 during foam delivery. (2) $R$ = 300 µm, $Q_g$ = 3.5 cm³/min, $Q_i$ = 1.2 cm³/h during foam generation, $\Delta t_i$ = 60 min, $Q_f$ = 3.5 cm³/min, $Q_i$ = 0 during foam delivery.